\documentclass[aps,twocolumn]{revtex4}
\usepackage{graphicx}
\usepackage{epsfig}
\begin{document}
\bibliographystyle{apsrev}
\title{Darwin Meets Einstein: LISA Data Analysis Using Genetic Algorithms}
\author{Jeff Crowder, Neil J. Cornish, and Lucas Reddinger}
\affiliation{Department of Physics, Montana State University, Bozeman, MT 59717}

\begin{abstract}

This work presents the first application of the method of Genetic Algorithms (GAs) to data
analysis for the Laser Interferometer Space Antenna (LISA). In the low frequency regime of
the LISA band there are expected to be tens of thousands galactic binary systems that will
be emitting gravitational waves detectable by LISA. The challenge of parameter extraction of
such a large number of sources in the LISA data stream requires a search method that can
efficiently explore the large parameter spaces involved. As signals of many of these sources
will overlap, a global search method is desired. GAs represent such a global search method for
parameter extraction of multiple overlapping sources in the LISA data stream. We find that GAs
are able to correctly extract source parameters for overlapping sources. Several optimizations
of a basic GA are presented with results derived from applications of the GA searches to
simulated LISA data.

\end{abstract}

\maketitle

\section{Introduction}

The Laser Interferometer Space Antenna (LISA)~\cite{PrePhaseA} is set to be launched in the
middle of the next decade. As LISA is an all-sky antenna, it will detect sources in all
directions, and across a great range of distances. The types of sources range from monochromatic
white dwarf binaries in our own galaxy to rapidly coalescing supermassive black hole binaries
in the distant reaches of the Universe. The challenge for analyzing the LISA data stream will
be pulling out the various parameters of as many of these sources as is possible. A large
impediment to completing this challenge is the many thousands of low frequency, effectively
monochromatic sources~\cite{evans,lip,hils,hils2,gils} that will be present in the LISA data
streams. Extracting the parameters from so many sources at once is analogous to determining
what every member in the audience of a rock concert is saying. As more sources overlap the
confusion grows rapidly~\cite{source_conf}. The name given to this issue is `The Cocktail Party
Problem' (see Ref.~\cite{MCMC} for a detailed discussion).

With so many sources, it will be impossible to extract the individual source parameters for
every source in the LISA band. This will leave a background of sources whose indeterminable
signals blend together into a confusion limited background. Several
studies~\cite{hils,hils2,gils,sterl,seth,bc} have indicated that the confusion noise
may dominate instrument noise at the low end of the LISA frequency range, so that other sources of
interest may be buried beneath the confusion background. For this reason a key goal of LISA
data analysis is to reduce the level of the confusion noise as much as possible.

Previous approaches to the extraction of parameters from the LISA data stream have used
several methods. Grid based template searches using optimal filtering provide a systematic
method to search through all possible combinations of gravitational wave sources, but the
computational cost of such a search appears to make it unfeasible~\cite{temp_grid}.
Other techniques applied to simulated LISA data involve iterative refinement of a
sequential search of sources~\cite{gclean,slicedice}, a tomographic approach~\cite{mohanty}, global iterative refinement, and ergodic exploration of the parameter space such as Markov Chain Monte Carlo (MCMC)
methods~\cite{MCMC}. At this time, however, it is not clear which of these techniques,
or which combination of techniques will provide the best solution to the Cocktail
Party Problem. 

Here we present the first application of the method of genetic algorithms~\cite{holland} to the challenge
of extracting parameters from a simulated LISA data stream containing multiple monochromatic
gravitational wave sources. The strength of this method lies in its searching capabilities,
and thus GAs might be used as the first step in dealing with the confusion background.
The initial solution could then be handed off to a MCMC algorithm~\cite{MCMC}, which
specializes in determining the nature of the posterior distribution function. 

In section~\ref{GA_search_section} we explore various factors that influence the performance
of a genetic search algorithm. A bare-bones algorithm is introduced in \ref{basics}, and
succeeding layers of complexity are added to this algorithm in \ref{mutation_subsection}
through \ref{amoeba_subsection}, with an emphasis on developing an efficient algorithm,
which is robust enough to handle the entire low frequency regime of the LISA detector.
Applications of the advanced algorithms to multiple source cases are shown
in \ref{multiple_source_subsection}. We conclude with a discussion of future improvements
and plans for the application of genetic algorithms to LISA data analysis.

\section{Genetic Search Algorithms}\label{GA_search_section}

The fundamental idea behind a genetic algorithm is the survival of the fittest.
It is because of this that genetic algorithms are often referred to as
evolutionary algorithms, though Darwin~\cite{origin} would probably have considered GAs
as ``Variation under Domestication'' since we are breeding toward a predetermined
goal. Through the process of continually evolving solutions to the given problem,
genetic algorithms provide a means to search the large parameter space that
we will be confronted with in the low frequency region of the LISA band.

A few definitions are in order before delving into our applications of
genetic algorithms to LISA data analysis. These definitions will refer to a hypothetical
search of the LISA data stream for $N$ monochromatic gravitational wave sources. The search will
take advantage of the F-statistic to reduce the search space to $3N$ parameters. The
hypothetical search will also involve the use of $n$ simultaneous, competing solution sets.\\
An {\bf organism }is a particular $3N$ parameter set that is a possible solution for the source parameters.\\
A {\bf gene} is an individual parameter within an organism.\\
A {\bf generation} is the set of all $n$ concurrent organisms.\\
{\bf Breeding} or cross-over is the process through which a new organism is formed from one or
more organisms of the previous generation.\\
{\bf Mutation} is a process which allows for variation of a organism
as it is bred from the organisms of the previous generation.\\
{\bf Elitism} is the technique of carrying over one or more of the best organisms
in one generation to the next generation.

A simplified genetic algorithm begins with a set of $n$ organisms that comprise the first generation.
The genes of this generation may be chosen at random or selected through some other process.
The organisms of each generation are checked for fitness, and those with the best fitness are
more likely to breed, with mutation, to form the organisms of the next generation.
With passing generations the organisms tend toward better solutions to the source parameters.
We use the F-statistic to measure the fitness of each organism.

\subsection{\label{basics}Basic Implementation}

For our investigations source frequencies were chosen to lie within the range
$f \in [0.999995,1.003164]$ {\rm mHz}. This range spans $100$ frequency bins of
width $\Delta f = 1/{\rm year}$. Amplitudes were restricted to the range $A \in [10^{-23},10^{-21}]$.
By use of the F-statistic our searches are reduced to frequency $f$, and sky location $\theta$ and $\phi$.
For a detailed description of the F-statistic and its use in reducing the search space
see Refs.~\cite{fstat,MCMC}.

A simple approach is to represent the values of each search parameter with binary strings.
The length of the strings determines the precision of the search, e.g. representing $\theta$
with a binary string of $8$ digits gives precision to $0.7^{\circ}$. Resolution is given by,
(parameter range)/$2^L$, where $L$ is the length of the binary string. Such a binary
representation allows for ease of mutation and breeding. We employed binary strings of length
$L=16$ for $f$, $L=13$ for $\theta$ and $L=14$ for $\phi$.

In this basic scheme, we first mutate the parent's parameter strings, and then breed the mutated gametes.
Simple mutation consists of flipping the binary digits of the parent's parameter strings with
probability PMR, the parameter mutation rate. A large PMR will tend to result in more variation in the
gametes, and thus the offspring, while a small PMR will lessen variation, resulting in more offspring
that resemble their parents.

We use a breeding pattern known as $1$-point crossover, which consists of the combination of
complimentary sections of the binary strings of two parent organisms. The cross-over point can
be chosen at random or fixed in advance. We chose a fixed cross-over with the cross-over
point occurring at the midpoint of the strings. As an example we show the breeding of a parameter
represented by strings that are $8$ digits long.

\begin{table}[hbtp]
\caption{Midpoint crossover for an 8 bit string}
\vspace*{0.1in}
\begin{tabular}{|c|rl|} \hline \hline
Parent $1$ & $0100$ & $1110$ \\
Parent $2$ & $0011$ & $0011$ \\
\hline
Offspring & $0011$ & $1110$\\
 \hline \hline
\end{tabular}
\label{breeding_example}
\end{table}

We will start with a basic search using $10$ organisms in each generation. The first generation has the
genes of its organisms chosen at random from their respective ranges. The probability of each of these
organisms being chosen for reproduction is proportional to its likelihood, $\cal L$ (known as fitness
proportionate cross-over). Mutated
gametes are formed using a PMR of $0.04$, and are bred using a single midpoint crossover.

\begin{figure}[h]
\includegraphics[angle=270,width=0.48\textwidth]{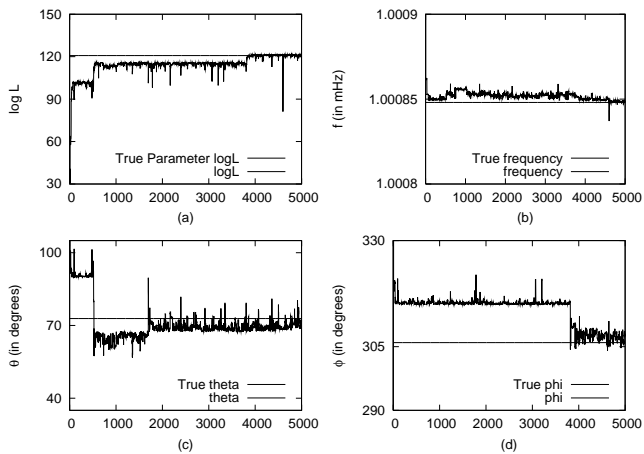}
\caption{\label{simple_scheme_plots}Basic Algorithm: Trace plots for (a) log likelihood, (b) frequency,
(c), $\theta$, and (d) $\phi$ for the basic implementation of a genetic algorithm. The y-axes
are the parameter values and log likelihoods of the best fit organism for each generation.
The x-axes are generation number.}
\end{figure}

Figure~\ref{simple_scheme_plots} shows trace plots of the log likelihood, frequency, $\theta$,
and $\phi$ for a source with ${\rm SNR}=15.4464$ and parameters: $A=1.97703 \times 10^{-22}$,
$f=1.000848032$ {\rm mHz},
$\theta=1.2713$, $\phi=5.34003$, $\iota=2.73836$, $\psi=1.43093$, and $\gamma_o=5.59719$
(it is this source that will be used repeatedly throughout the paper). The plotted values were
for the organism with the best fit in each generation. As can be seen the parameters are well
determined with even this basic scheme, though the noise in the data stream pushes them off
their true values. The parameter values are shifted by $\delta f = -1.5 \times 10^{-9}$ Hz,
$\delta \theta = 2.9^\circ$ and $\delta \phi = -1.5^\circ$ from their input values. These
shifts are consistent with the error predictions from a Fisher matrix analysis:
$\Delta f = 1.7 \times 10^{-9}$ Hz, $\Delta \theta = 3.5^\circ$ and $\Delta \phi = 1.9^\circ$.
The cost of the search is measured in terms of the number of calls to the
F-statistic routine and is given by $\$ = n\times g$, where $g$ is the generation number.
Typical runs of our basic genetic algorithm cost $\$ = 32650$ calls. This should be compared to a
grid based search across
the same frequency range, which, for a minimal match of ${\rm MM}=0.9$, would require $\$=110,000$
calls to the F-statistic routine (this value is $2^{3/2}$ larger than that quoted in Ref.~\cite{MCMC}
as our earlier calculations used a noise level that was $\sqrt{2}$ larger than the LISA
baseline due to a mix up between one and two sided noise spectral densities).

While the basic algorithm is sufficient for finding a solution, it is not efficient.
Next we will discuss adjustments to the algorithm that will improve its efficiency, and make
it considerably cheaper than a grid based search.

\subsection{\label{mutation_subsection}Aspects of Mutation}

\begin{figure}[h]
\includegraphics[angle=270,width=0.48\textwidth]{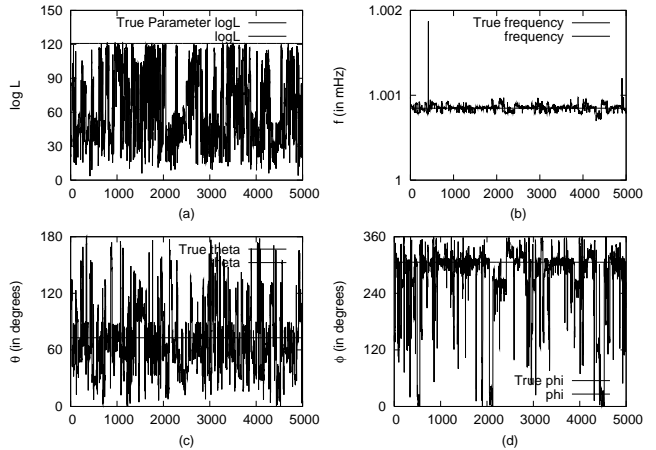}
\caption{\label{lg_PMR_plots}Large Mutation Rate: Trace plots for (a) log likelihood, (b) frequency,
(c), $\theta$, and (d) $\phi$ for the basic implementation of a genetic algorithm with ${\rm PMR}=0.1$.
The y-axes are the parameter values and log likelihoods of the best fit organism for each generation.
The x-axes are generation number.}
\end{figure}

\begin{figure}[h]
\includegraphics[angle=270,width=0.48\textwidth]{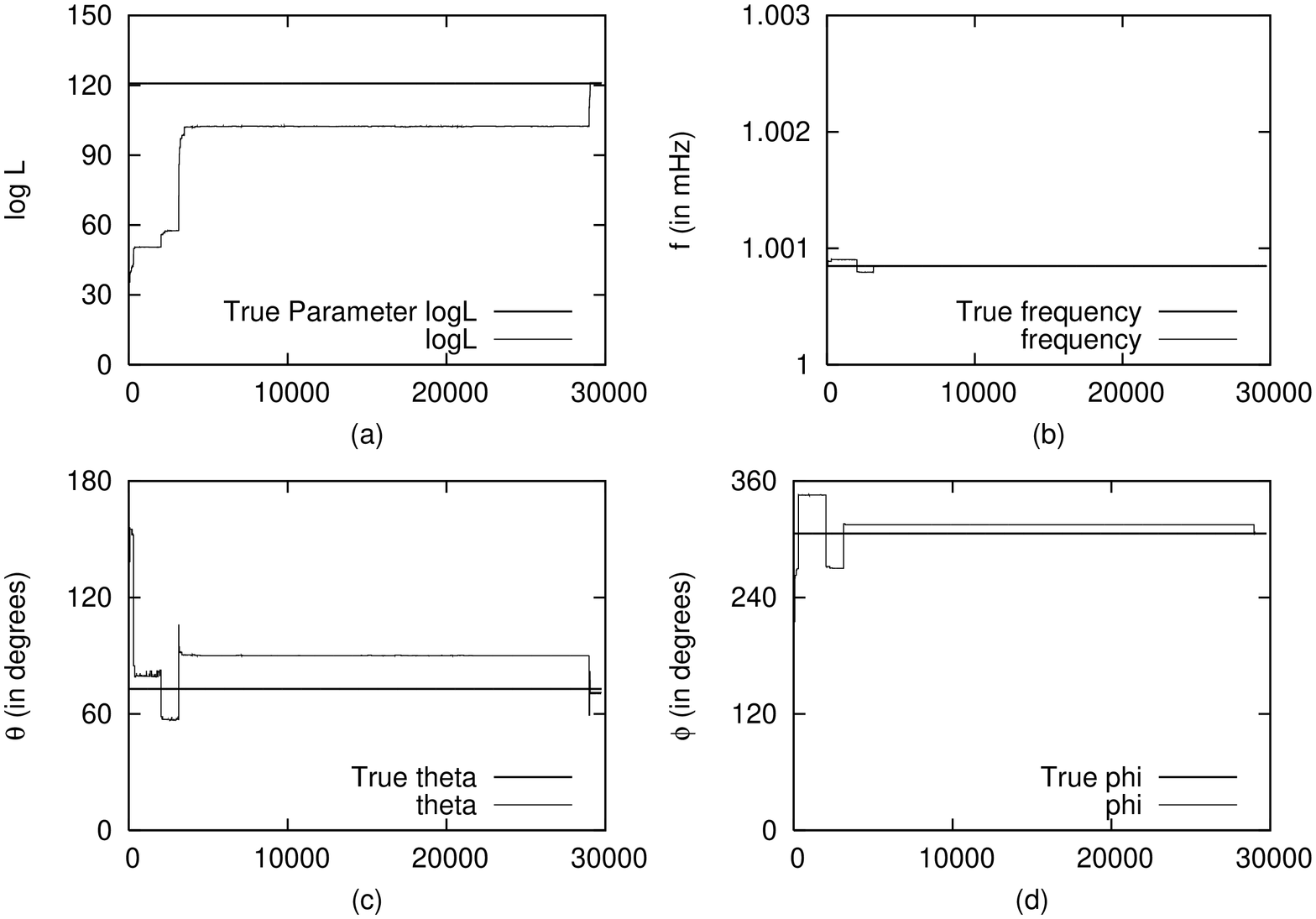}
\caption{\label{sm_PMR_plots}Small Mutation Rate: Trace plots for (a) log likelihood, (b) frequency,
(c), $\theta$, and (d) $\phi$ for the basic implementation of a genetic algorithm with ${\rm PMR}=0.001$.
The y-axes are the parameter values and log likelihoods of the best fit organism for each generation.
The x-axes are generation number.}
\end{figure}

In the previous example the PMR was set at the fairly low value of $0.04$.
Figure~\ref{lg_PMR_plots} shows trace plots for the same search, but with ${\rm PMR}=0.1$.
While the ${\rm PMR}=0.04$ example shows a tendency for small deviations from the improving solutions,
the larger PMR search allows large swings in the solution away from a good fit to the true source parameters.
On the other hand, Figure~\ref{sm_PMR_plots} shows how a small PMR ($0.001$) can cause the rate of progress
to be greatly slowed. A small mutation rate slows the exploration of the likelihood surface.

\begin{figure}[h]
\includegraphics[angle=270,width=0.48\textwidth]{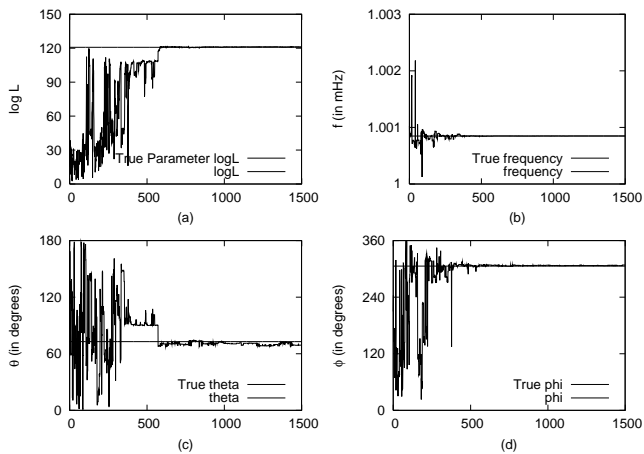}
\caption{\label{gen_sim_anneal_plots}Genetic Simulated Annealing: Trace plots for (a) log likelihood,
(b) frequency, (c), $\theta$, and (d) $\phi$ for the basic implementation of a genetic algorithm with
the inclusion of genetic simulated annealing. The y-axes are the parameter values and log likelihoods
of the best fit organism for each generation. The x-axes are generation number.}
\end{figure}

As these examples show, choosing the proper PMR can have a significant effect on the efficiency of the
algorithm. Knowing which value is the proper choice a priori is impossible. Furthermore, at different
phases of the search, different values of the PMR will be more efficient than those same values
at other phases. Early on in the search a large PMR is desirable for increased exploration. Once convergence
to the solution has begun, a smaller PMR is preferable, to prevent suddenly mutating away from the solution.
One can imagine a process which changes the PMR in a manner analogous to the simulated annealing process,
where we start the PMR high (hot) and lower (cool) it in succeeding generations. In fact, this process
in sometimes called simulated annealing in the GA literature. Figure~\ref{gen_sim_anneal_plots}
shows trace plots for the same source, using a genetic (PMR) simulated annealing scheme given by:

\begin{equation}
{\rm PMR} = \left\{
\begin{tabular}{ll}
${\rm PMR}_{\rm f} \left(\frac{PMR_{\rm i}}{PMR_{\rm f}}\right)^{\frac{g_{\rm cool}-g}{g_{\rm cool}}}$
& $0 < g < g_{\rm cool}$ \\
${\rm PMR}_{\rm f}$ &  $g \geq g_{\rm cool} $
\end{tabular}
\right.
\end{equation}
where ${\rm PMR}_{\rm i}=0.2$, ${\rm PMR}_{\rm f}=0.01$, $g$ is the generation number,
and $g_{\rm cool}=1000$ is the
last generation of the cooling process. The best choice of values for this scheme is again impossible
to know a priori. In section~\ref{gg_subsection} we will see how ``Genetic Genetic Algorithms'' are able
to provide a natural solution to this problem.

\subsection{\label{cost_subsection}The effect of the number of organisms on efficiency}

While choosing the PMR is one degree of freedom in our basic schema, another is the number of
organisms used in the search. Here we look at how the choice of the number of organisms effects the
efficiency of the algorithm. The efficiency is inversely related to the computational cost \$, which
is measured by the number of calls to the function calculating the F-statistic (where the bulk of
calculations for an organism are performed), which occurs once per newly formed organism. For example,
in Figure~\ref{simple_scheme_plots} there are $10$ organisms in the search and the
search surpasses the true parameter log likelihood value at $3851$ generations. Thus its
computational cost is $\$=38510$ (function calls). 

\begin{figure}[h]
\includegraphics[angle=270,width=0.48\textwidth]{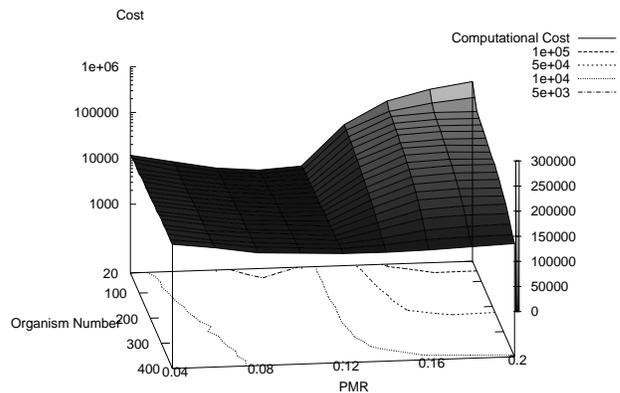}
\caption{\label{3d_cost_plot}Average Computational Cost as a function of PMR and the number of organisms.
The z-axes is the average computational cost calculated from $1000$ searches.}
\end{figure}

The data in Figure~\ref{3d_cost_plot} shows the interplay of the number of organisms with the
PMR (held constant within each data run) and their effects on the computational cost. We would
expect that relatively large PMRs would be less efficient as was seen in
subsection~\ref{mutation_subsection} (and will show up in Figure~\ref{elitism_pmr_cost_plots}).
The size of the effect, however, is modified by the number of organisms in the search. For example,
one can find from Figure~\ref{3d_cost_plot} that the minimum cost ($\$ = 4492$) for a $20$ organism
search occurs when ${\rm PMR}=0.1$, however for $400$ organisms in the search the minimum cost
($\$ = 7490$) is at ${\rm PMR}=0.14$.

The addition of more organisms in the search provides a kind of stability to the system that decreases
the chances of mutating away from good solutions. With just a handful of organisms, and a large PMR,
the chances are higher of each organism undergoing a large mutation in at least one parameter. However,
with hundreds of organisms the probability of all organisms undergoing such a mutation drops appreciably.
Then in the succeeding generation, those organisms that remained a good fit are much more likely to breed
the offspring of the next generation. However, this does not hinder great leaps forward. To illustrate
this point we will use the data shown in Figure~\ref{simple_scheme_plots}. In going from the $7^{th}$
to the $8^{th}$ generation the value of the likelihood of the best fit organism jumps from
$1.48\times10^{13}$ to $6.02\times10^{20}$. As the probability of breeding is set by the value of the
organism likelihood, that new best fit organism is going to be the primary breeder of the next
generation (though it is possible that a second organism has also jumped to a point in parameter
space with a similar likelihood value).

Increasing the number of organisms not only provides this stabilizing effect, it also provides more chances
per generation for improvements due to mutations. One cannot, however, simply throw more organisms at the
problem without paying a price; that price will be an eventual drop in efficiency. As an extreme example,
imagine using the basic scheme describe in \ref{basics} and putting $40000$ organisms into the search.
Even if one of the randomly chosen organisms matched the best fit parameters, the computational cost
($\$=40000$) is already larger than the cost of using $10$ organisms ($\$= 38510$).
Figure~\ref{3d_cost_plot} provides a snapshot of the how this choice effects efficiency.

\subsection{\label{elitism_subsection}Elitism}

Elitism is akin to cloning. It allows for a perfect copy of an organism or organisms to be bred
into the next generation. Including elitism is another way to provide a stabilizing force across
generations. This allows for a larger PMR to enhance exploration without the danger of moving off
the best fit solution.

\begin{figure}[h]
\includegraphics[angle=270,width=0.48\textwidth]{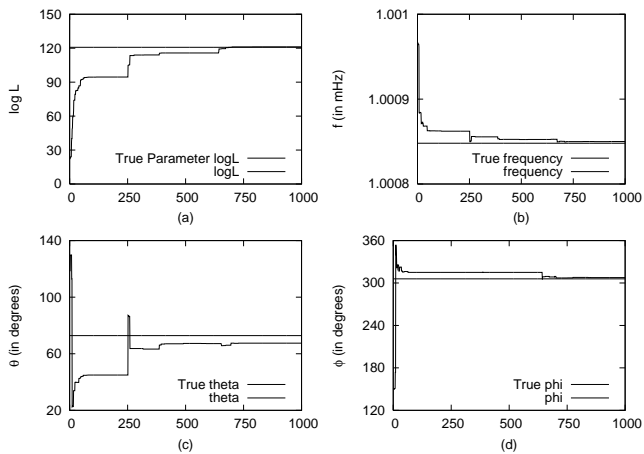}
\caption{\label{simple_scheme_w_elitism_plots}Elitism: Trace plots for (a) log likelihood,
(b) frequency, (c), $\theta$, and (d) $\phi$ for the basic implementation of a genetic
algorithm with ${\rm PMR}=0.1$ and single organism elitism. The y-axes are the parameter values
and log likelihoods of the best fit organism for each generation. The x-axes are generation number.}
\end{figure}

Figure~\ref{simple_scheme_w_elitism_plots} shows trace plots for the nominal source with ${\rm PMR}=0.1$
and a single elite organism being cloned at each generation. As expected there is increased
exploration (compared to results shown in Figure~\ref{simple_scheme_plots}) due to the larger PMR,
but unlike the results shown in Figure~\ref{lg_PMR_plots}, convergence is now helped by the cloned
organism.

\begin{figure}[h]
\includegraphics[angle=270,width=0.48\textwidth]{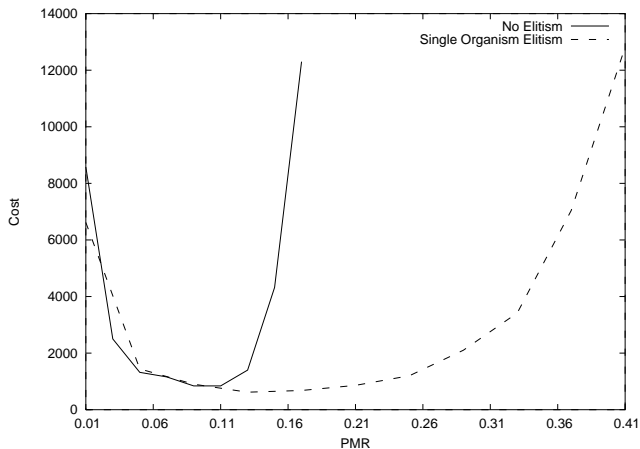}
\caption{\label{elitism_pmr_cost_plots}Average Computational Cost for no elitism and
single organism elitism. Data points are determined by the average of $100$ distinct searches.}
\end{figure}

Figure~\ref{elitism_pmr_cost_plots} shows a plot relating the average computational cost to the PMR
for the case of no elitism, and the case where a single organism is cloned. Computational cost is now
derived from the average number of newly formed organisms (note: a cloned organism does not increase
computational cost, as all of its associated values are already known). The plot shows the average
computational cost of $100$ searches, using $20$ organisms, of a given source
(${\rm SNR}=19.2335$ and parameters: $A=1.61486e-22$, $f=1.003$ {\rm mHz}, $\theta=0.8$, $\phi=2.14$,
$\iota=0.93245$, $\psi=2.24587$, and $\gamma_o=5.29165$). As was expected, elitism has allowed for
a larger PMR, compared to the zero elitism case, increasing the parameter space exploration
without sacrificing efficiency.

If one decides to use elitism there is the additional choice of how many elite organisms will be
cloned at each generation. At one extreme all organisms are cloned, in which case there
is no exploration beyond the first generation. At the other extreme of no elitism the algorithm
is unstable against large PMR values, as was seen in Figure~\ref{lg_PMR_plots}. There is a balance
to be struck between the amount of elitism and the size of the PMR that will provide the most
efficient scheme, but the exact nature of the balance can depend on the nature of the search.
We describe a solution to this problem in \ref{gg_subsection}.

\subsection{\label{annealing_subsection}Simulated Annealing}

Simulated annealing is a technique that effectively makes the detector more noisy, thus lessing the
range of the likelihood function. This increases the probability of choosing poorer sources for
reproduction, which allows for a more thorough exploration of the likelihood surface.
Think of the likelihood as a partition function $Z = C \exp(-\beta E)$, in which the role of the
energy is played by the log likelihood, $E = (s-h \vert s-h)$, and $\beta$ plays the role of the
inverse temperature. Heating up the system (lowering $\beta$) lowers the likelihood range,
providing for increased exploration. Starting hot, we use a power law cooling schedule given by:

\begin{equation}
\beta = \left\{
\begin{tabular}{ll}
$\beta_0 \left(\frac{1}{2\beta_0}\right)^{g/g_{\rm cool}}$ & $0 < g < g_{\rm cool}$ \\
$\frac{1}{2}$ &  $g \geq g_{\rm cool} $
\end{tabular}
\right.
\end{equation}
where $\beta_0$ is the initial value of the inverse temperature, $g$ is the generation number,
and $g_{\rm cool}$ is the last generation of the cooling process (subsequent generations have $\beta=1/2$).
As the likelihood is a sharply peaked function, we found for a single source an initial value of
$\beta_0 \sim 1/100$ was sufficient to speed the process. For multiple source searches increasing
that by factors of $3$ to $5$ produced more efficient explorations. Similarly, for multiple sources
an increase in $g_{\rm cool}$ was needed to properly explore the surface. This increase scaled
roughly linearly with the number of sources.

This mode of simulated annealing, which will be referred to as standard simulated annealing, is markedly
different than the genetic version of simulated annealing discussed in \ref{mutation_subsection}. Standard
simulated annealing alters the search space, using the heat/energy to smooth the likelihood surface,
whereas in genetic simulated annealing the search space was left unchanged and the heat/energy of the
organisms was increased via the larger PMRs.

\begin{figure}[h]
\includegraphics[angle=270,width=0.48\textwidth]{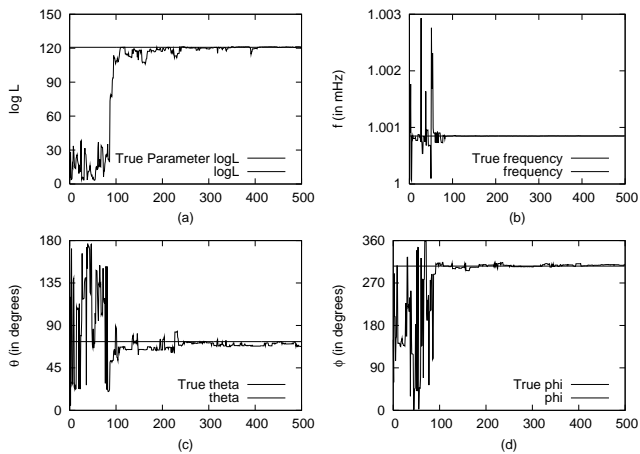}
\caption{\label{simple_scheme_standard_annealing_plots}Standard Simulated Annealing: Trace plots for
(a) log likelihood, (b) frequency, (c), $\theta$, and (d) $\phi$ for the basic implementation of a
genetic algorithm with the inclusion of standard simulated annealing and ${\rm PMR}=0.04$. The y-axes are
the parameter values and log likelihoods of the best fit organism for each generation. The x-axes are
generation number.}
\end{figure}

Figure~\ref{simple_scheme_standard_annealing_plots} shows trace plots of the log likelihood,
frequency, $\theta$, and $\phi$ searching for the same source as in Figure~\ref{gen_sim_anneal_plots}.
The only change between the two examples is the type of annealing process. For this
run ${\rm PMR}=0.04$, $\beta_0=1/100$, and $g_{\rm cool}=300$.

\subsection{\label{gg_subsection}Giving more control to the algorithm}

In the previous examples, choices were required as to what PMR or which degree of elitism should be used
with a particular source to provide the most efficient search. In making those choices, we are searching
for a solution that depends on the information in the data stream. Just as we use the power of the
genetic algorithm to search for the parameters of the gravitational wave sources that contribute to the
data stream, we can also use that same power to search for efficient values for PMR or elitism. 

Treating the PMR, elitism, or other factors in the genetic algorithm like a source parameter these
factors can be elevated, or one might say demoted, to the same level as the source parameters. We
mentioned this at then end of subsection~\ref{mutation_subsection} and have implemented this idea
for the PMR. The initial PMR for each organism is chosen randomly, and the PMR for each organism in
the next generation is bred just as $f$, $\theta$, and $\phi$ are, based on organism fitness. This
changes the nature of the algorithm from a simple genetic algorithm to a genetic-genetic algorithm (GGA),
in which a factor, or factors, determining the search for the source parameters evolve along with the
organisms.

\begin{figure}[h]
\includegraphics[angle=270,width=0.48\textwidth]{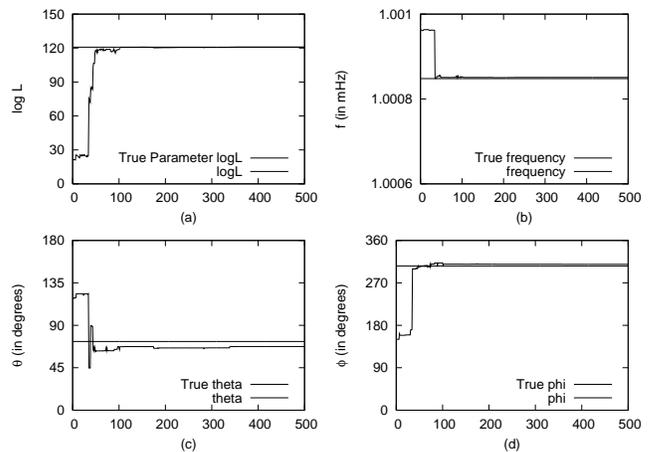}
\caption{\label{gga_PMR_plots}Genetic-Genetic Algorithm: Trace plots for (a) log likelihood,
(b) frequency, (c), $\theta$, and (d) $\phi$ for a genetic-genetic algorithm in which the PMR evolves
with the organisms. The y-axes are the parameter values and log likelihoods of the best fit organism for
each generation. The x-axes are generation number.}
\end{figure}

\begin{figure}[h]
\includegraphics[angle=270,width=0.48\textwidth]{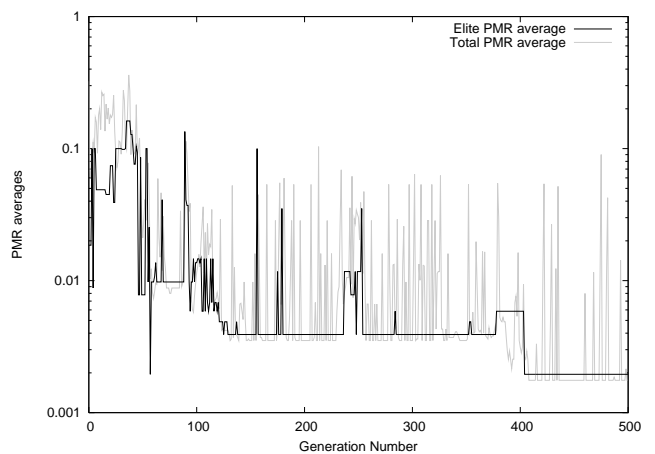}
\caption{\label{gga_PMR_schedule_plots}Genetic-Genetic Simulated Annealing of the PMR: Trace plots for
the PMR as it evolves with the organisms. The data for this plot is from the same run that produced the
data in Figure~\ref{gga_PMR_plots}.}
\end{figure}

Figure~\ref{gga_PMR_plots} shows trace plots for a GGA with the PMR evolving with the organisms. This run
includes the simulated annealing scheme used in the previous example and elitism of the single best fit
organism. Figure~\ref{gga_PMR_schedule_plots} shows the evolution of the PMR for the same run.
The `genetic simulated annealing' scheme is visible in the plot with the larger PMRs more efficient
earlier on, and smaller PMRs dominating in the later stages. As the evolving PMR values range over nearly
two orders of magnitude, it is easy to see why a single, constant choice for the PMR would be so much
less efficient. Also, as one can see from the data presented, the variations in the frequency are
significantly smaller than those of $\theta$ and $\phi$. We can extend the idea of tailored PMRs
beyond the organism, and down to the gene. Giving a separate PMR to each parameter will allow for
even better adaptation. (In the natural world organisms control their mutation rates by building
in DNA repair mechanisms to counteract the externally determined mutation rate set by cosmic rays
and other pathogens).

\subsection{\label{multiple_source_subsection}Multiple sources in the data stream}

At the low end of the LISA band there will be many thousands of sources. Thus, we expect to see
multiple sources even in small segments of the data stream such as the one we have been
considering. Simulations point to bright source densities of up to one source per five modulation
frequency bins ($f_{\rm mod} = 1/{\rm year}$)~\cite{seth}. Thus, any search algorithm must be
able to perform multiple source searches at the low end of the LISA band.

\begin{figure}[h]
\includegraphics[angle=270,width=0.48\textwidth]{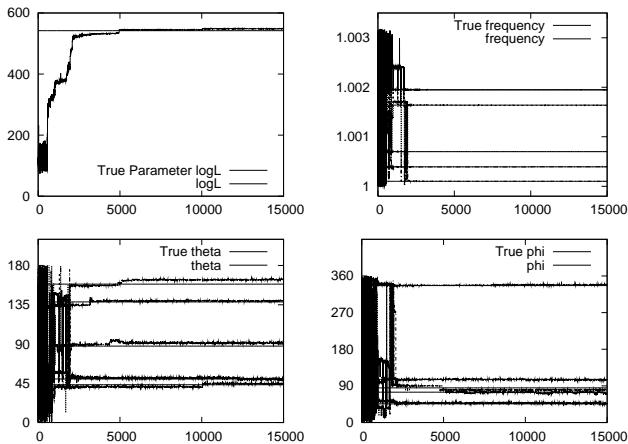}
\caption{\label{five_source_plots}Genetic algorithm search for 5 sources: Trace plots for (a) log likelihood,
(b) frequency, (c), $\theta$, and (d) $\phi$ for a genetic algorithm searching for the presence of
five gravitational wave sources in the data stream. The y-axes are the parameter values and log likelihoods
of the best fit organism for each generation. The x-axes are generation number.}
\end{figure}

Figure~\ref{five_source_plots} shows an implementation of the GGA with standard simulated annealing
to a LISA data stream snippet of width $100 f_{\rm mod}$, containing five monochromatic binary systems.
The standard simulated annealing was completed in the first $g_{\rm cool}=4000$ generations, by
which time the GGA had separated out the values for the source frequencies and co-latitudes.
The grouping of azimuthal angles was separated soon thereafter, with minor modifications of the
parameters occurring over the next $5000$ generations. Search results are summarized in
Table~\ref{five_source_table}. The GGA accurately recovered the source parameters in this
and similar multiple ($3-5$) source data sets, converging to a best fit solution in less
than $5000$ generations per source with $10$ organisms per generation, so long as the
source correlation coefficients were below $\sim 0.25$. The intrinsic parameters for the sources
were recovered to within $2\sigma$ of the true parameters (based on a Fisher Information Matrix
estimate of the uncertainties of the recovered parameters). When highly correlated sources are used,
the GGA spends a correspondingly longer time to pick out the source parameters. Investigations in
this area were limited. A full study of the affect of source correlation on computational cost is
to be carried out in the future.

\begin{table}[t]
\caption{GGA search for 5 galactic binaries. The frequencies are quoted relative to 1 {\rm mHz}
as $f = 1 \, {\rm mHz} + \delta f$ with $\delta f$ in $\mu {\rm Hz}$. All angles are quoted in radians.}
\begin{tabular}{|l|c|c|c|c|c|c|c|c|}
\hline
  & SNR & $A$ ($10^{-22}$) & $\delta f $ & $\theta$ & $\phi$ & $\psi$ & $\iota$ & $\varphi_0$ \\
 \hline 
True 	& 12.7	& 1.02	& 1.638	& 2.77	& 1.48	& 2.28	& 0.886	& 0.273 \\
GA ML 	& 11.6 	& 1.08 	& 1.635 & 2.86 	& 1.40 	& 2.63	& 1.02 	& 5.94	\\
 \hline 
True 	& 19.3	& 2.23	&0.7000	& 2.41	& 5.87	& 0.435	& 1.88	& 4.29 \\
GA ML 	& 17.7 	& 2.11 	&0.7008	& 2.43 	& 5.90 	& 0.460	& 1.86	& 4.20	\\
 \hline 
True 	& 17.8	& 1.74	&0.3937	& 0.756	& 1.85	& 1.41	& 2.02	& 3.09 \\
GA ML 	& 17.0 	& 1.80 	&0.3942	& 0.777	& 1.84 	& 1.27	& 1.95 	& 2.57	\\
 \hline 
True 	& 15.8	& 2.16 	&1.002	& 1.53	& 1.30	& 1.35	& 1.70	& 4.63 \\
GA ML 	& 14.8 	& 2.17	&1.002 	& 1.59 	& 1.28 	& 1.37	& 1.68 	& 4.68	\\
 \hline 
True 	& 12.1	& 0.836	& 1.944	& 0.872	& 0.802	& 1.56	& 0.805	& 3.87 \\
GA ML 	& 11.8 	& 1.09 	& 1.950	& 0.876	& 0.803	& 2.87 	& 1.09 	& 3.48	\\
\hline 
\end{tabular}
\label{five_source_table}
\end{table}

\subsection{\label{amoeba_subsection}Using Active Organisms}

So far all of the organisms that have been discussed are passive organisms. They are passive in the
sense that once they are bred, the organisms themselves remain unchanged,
and are simply used to breed the next generation. One can imagine organisms that `learn' during their
lifetime, advancing toward a better solution. Directed search methods such as an uphill simplex, i.e.
an amoeba, provide a means for organisms to advance within a generation. As the likelihood surface is
not entirely smooth, the simplex may get stuck in a local maximum that is removed from the global maximum.
So the generational process is still necessary to ensure full exploration of the surface.
One approach is to use the
the parameters bred from one generation as the centroid of the simplex (amoeba), which will
then proceed to move uphill across the likelihood surface. Another approach, that we will describe
in a future publication, is to use `Genetic Amoeba', where genes code for each vertex of the simplex.
The amoeba are allowed to breed after they have found enough food (i.e. increased their likelihood
by a specified amount). Amoeba that eat well get to breed the most often and have the most
offspring.

\begin{figure}[h]
\includegraphics[angle=270,width=0.48\textwidth]{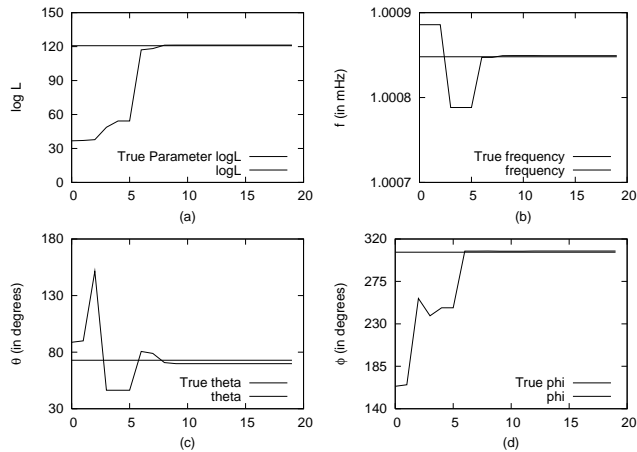}
\caption{\label{amoeba_inclusion_plots}GGA with a directed organism: Trace plots for (a) log likelihood,
(b) frequency, (c), $\theta$, and (d) $\phi$ for a GGA with a single directed organism. The y-axes are
the parameter values and log likelihoods of the best fit organism for each generation. The x-axes are
generation number.}
\end{figure}

Figure~\ref{amoeba_inclusion_plots} shows trace plots for an implementation of a GGA with a
single directed organism per generation. The other $9$ organisms were the standard passive organisms.
There was elitism with a single organism being cloned into the succeeding generation, and there was no
standard simulated annealing. What is missing from the plot is the computational cost. While computational
cost can easily be derived from the plots with passive organisms, active organisms, such as an uphill
simplex involve multiple calls to the F-statistic function within a single generation. At the
$8^{th}$ generation, where the search surpasses the true likelihood value, the computational cost is
$\$=876$. This cost is slightly lower than the cost of a GGA with only passive organisms at the point
where its search surpasses the likelihood value for the true parameters. However, for true LISA data,
we will not know the true parameters, and thus will have to allow the algorithms to undergo extended
runs to ensure they have fully explored the space and found the global maximum. The higher computational
cost per generation of the simplex method (which averages $\sim 100$ calls to find a local maximum)
will quickly lead to a higher total cost of the search. Other directed methods that are more efficient
than an uphill simplex may provide an alternative that will provide an overall improvement in
efficiency. Future work will include an examination of other possibly more efficient directed methods,
and a detailed study of the Genetic Amoeba algorithm.

\section{Conclusions}\label{conclude}

This work is the first application of a genetic algorithm to the search of gravitational wave source
parameters. We have shown that the method is a feasible search method capable of handling multiple sources
in a restricted frequency range. Next we will seek to determine the limits of the algorithm both in terms of
source number and source density across the low frequency regime of the LISA band. While an optimal
solution would employ a matched filter that includes every resolvable source in the LISA band~\cite{MCMC},
it is unlikely that a direct search for this ``super template'' is the best way to proceed. A better
approach may be to start with a collection of ``single cell'' organism that each code for a
single source (or possibly small collections of highly correlated sources), then combine these
cells into a multi-cellular organism that searches for the super template. This approach is motivated
by the {\em cellular slime molds} Dictyostelida and Acrasida, which spend most of their lives as
separate single-celled amoeboid protists, but upon the release of a chemical signal, the individual
cells aggregate into a great swarm that acts as a single multi-celluar organism, capable of movement
and the formation of large fruiting bodies. Future work will also
include investigations into algorithm optimization and adaptation of the algorithm to other source
types (e.g. coalescing binaries).  Furthermore, a thorough study comparing the computational cost and
resolution capabilities of an optimized genetic algorithm to other (optimized) search methods like
Markov Chain Monte Carlo searches, gClean, Slice \& Dice, and Maximum Entropy methods would provide
guidance on how to proceed in solving the LISA Data Analysis Challenge. \\

\section*{Acknowledgements}
This work was supported by NASA Grant NNG05GI69G and NASA Cooperative Agreement NCC5-579.


\begin{thebibliography}{99}

\bibitem{PrePhaseA} P.~L. Bender, et al., \textit{LISA Pre-Phase A Report; Second Edition}, MPQ 233 (1998).
\bibitem{evans} C. R. Evans, I. Iben \& L. Smarr, ApJ {\bf 323}, 129 (1987).
\bibitem{lip} V. M. Lipunov, K. A. Postnov \& M. E. Prokhorov, A\&A {\bf 176}, L1 (1987).
\bibitem{hils} D. Hils, P. L. Bender \& R. F. Webbink, ApJ {\bf 360}, 75 (1990).
\bibitem{hils2} D. Hils \& P. L. Bender, ApJ {\bf 537}, 334 (2000).
\bibitem{gils} G. Nelemans, L. R. Yungelson \& S. F. Portegies Zwart, A\&A {\bf 375}, 890 (2001).
\bibitem{source_conf} J. Crowder \& N.J. Cornish, Phys. Rev. D{\bf 70}, 082005 (2004).
\bibitem{MCMC} N.J. Cornish \& J. Crowder, Phys. Rev. D{\bf 72}, 043005 (2005).
\bibitem{sterl} A. J. Farmer \& E. S. Phinney, Mon. Not. Roy. Astron. Soc. {\bf 346}, 1197 (2003).
\bibitem{seth} S. Timpano, L. J. Rubbo \& N. J. Cornish, gr-qc/0504071 (2005).
\bibitem{bc}  L. Barack \& C. Cutler, Phys. Rev. D{\bf 70}, 122002 (2004).
\bibitem{temp_grid} J. R. Gair, L. Barack, T. Creighton, C. Cutler, S. L. Larson, E. S. Phinney \&
M. Vallisneri, Class. Quant. Grav. {\bf 21}, S1595 (2004).
\bibitem{gclean} N.J. Cornish \& S.L. Larson, Phys. Rev. D{\bf 67}, 103001 (2003).
\bibitem{slicedice}  N.J. Cornish, {\em Talk given at GR17, Dublin, July (2004)}; N.J. Cornish,
L.J. Rubbo \& R. Hellings, {\em in preparation} (2005).
\bibitem{mohanty} M. S. Mohanty, \& R. K. Nayak, gr-qc/0512014 (2005).
\bibitem{holland} J. Holland, {\em Adaptation in Natural and Artificial Systems}, (Ann Arbor, Michigan,
University of Michigan Press, 1975).
\bibitem{origin} C. Darwin, {\em The Origin of Species}, (J. Murray, London, 1859).
\bibitem{fstat} P. Jaranowski, A. Krolak \& B. F. Schutz, Phys. Rev. D{\bf 58} 063001 (1998).


Mohanty and Nayak, gr-qc/0512014

\end{thebibliography}
\end{document}